\documentclass[aps,preprint]{revtex4}%
\usepackage{amsfonts}
\usepackage{amsmath}
\usepackage{amssymb}
\usepackage{graphicx}%
\begin{document}
\title{Hawking radiation for higher dimensional Einstein-Yang-Mills linear dilaton
black holes }
\author{S. Habib Mazharimousavi$^{\ast}$}
\author{I. Sakalli$^{\ddag}$}
\author{M. Halilsoy$^{\dag}$}
\affiliation{Department of Physics, Eastern Mediterranean University,}
\affiliation{G. Magusa, north Cyprus, Mersin-10, Turkey}
\affiliation{$^{\ast}$habib.mazhari@emu.edu.tr}
\affiliation{$^{\dagger}$izzet.sakalli@emu.edu.tr}
\affiliation{$^{\ddag}$mustafa.halilsoy@emu.edu.tr}

\begin{abstract}
Recently, Hawking radiation from a $4-$dimensional linear dilaton black hole
solution to Einstein-Maxwell-Dilaton (EMD) theory, via a method of computing
exactly the semi-classical radiation, has been derived by Cl\'{e}ment, et al.
Their results show that, whenever solution is available to the massless scalar
wave equation, an exact computation of the radiation spectrum leads to the
Hawking temperature $T_{H}$. We apply the same method to derive Hawking
radiation spectrum for higher dimensional linear dilaton black holes in the
Einstein-Yang-Mills-Dilaton (EYMD) and Einstein-Maxwell-Dilaton (EMD)
theories. Our results show that the radiation with high frequencies for these
massive black holes reveal some remarkable information about the $T_{H}$.

\end{abstract}
\maketitle

\section{Introduction}

Although today there are different methods to compute the Hawking radiation,
(see for instance \cite{R1,R2,R3,R4,R5,R6}), it still remains of interest to
consider alternative derivations. On the other hand, none of them is
completely conclusive. Nevertheless, the most direct is Hawking's original
study \cite{R1,R2}, which computes the Bogoliubov coefficients between in and
out states for a realistic collapsing black hole. The most significant remark
on this study is that a black hole can emit particles from its event horizon
with a temperature proportional to its surface gravity. Another elegant
contribution was made to the Hawking radiation by Unruh \cite{R7}. He showed
that it is possible to obtain the same Hawking temperature when the collapse
is replaced by appropriate boundary conditions on the horizon of an eternal
black hole. Instead of computing the Bogoliubov coefficients in order to
obtain the black hole radiation, one may alternatively compute the reflection
and transmission coefficients of an incident wave by the black hole. This
method works best if the wave equation can be solved, exactly. From now on, we
designate this method with "semi-classical radiation spectrum method" and
abbreviate it as \textit{SCRSM}.

Recently, Cl\'{e}ment, et al \cite{R8} have studied the \textit{SCRSM} for a
class of non-asymptotically flat charged massive linear dilaton black holes.
The metric of the associated linear dilaton black holes is a solution to the
Einstein-Maxwell-Dilaton (EMD) theory in 4-dimensions. It is shown that in the
high frequency region, the \textit{SCRSM} for massive black holes yield the
same temperature with the surface gravity method. Their result for a massless
black hole is in agreement with the fact that a massless object can't radiate.

In this Letter, we shall follow \cite{R8} as a guide and extend the
application of \textit{SCRSM} to $N$-dimensional linear dilaton black holes,
which are the solutions to Einstein-Yang-Mills-Dilaton (EYMD) theory
\cite{R9}. The spacetimes describing these linear dilaton black holes are
non-asymptotically flat. First, we consider the statistical Hawking
temperature of the massive linear dilaton black holes, computed by using the
surface gravity and discuss their evaporation process. According to the
Stefan's law, we show that only 4-dimensional black holes evaporate in an
infinite time, while the others $N\geqslant5$\ can evaporate in a finite time.
In both cases, during the evaporation process, the Hawking temperature remains
constant for a given dimension. Besides this, the constant value of the
Hawking temperature increases with the dimensionality number. We then apply
the \textit{SCRSM} to the massive linear dilaton black holes and show that
this computation exactly matches with the statistical Hawking temperature in
the high frequency region. We explain how the massless extreme black holes do
not radiate by making a connection between our work and \cite{R8}.

In Sec. II we present and solve the EYMD equations and by using the massless
Klein-Gordon equation, radiation spectrum is derived. In Sec. III we repeat
similar analysis for the EMD theory, intending to compare with the EYMD
results. The Letter ends with Conclusion in Sec. IV.

\section{Einstein-Yang-Mills-Dilaton (EYMD) black hole radiation spectrum}

Recently, we obtained a static spherically symmetric higher dimensional linear
dilaton black hole solution in EYMD theory \cite{R9}. The metric ansatz is
given by
\begin{equation}
ds^{2}=-f\left(  r\right)  dt^{2}+\frac{dr^{2}}{f\left(  r\right)  }+h\left(
r\right)  ^{2}d\Omega_{N-2}^{2}%
\end{equation}
where $f\left(  r\right)  $ and $h\left(  r\right)  $ are only functions of
$r$ and the spherical line element is
\begin{equation}
d\Omega_{N-2}^{2}=d\theta_{1}^{2}+\underset{i=2}{\overset{N-3}{%
{\textstyle\sum}
}}\underset{j=1}{\overset{i-1}{%
{\textstyle\prod}
}}\sin^{2}\theta_{j}\;d\theta_{i}^{2}%
\end{equation}
in which $0\leq\theta_{k}\leq\pi$ with $k=1..N-3$, and $0\leq\theta_{N-2}%
\leq2\pi.$ Our action is
\begin{align}
\mathbf{I}  &  =-\frac{1}{2}\int\nolimits_{\mathcal{M}}d^{N}x\sqrt{g}\left[
R-\left(  \mathbf{\nabla}\Phi\right)  ^{2}-e^{2\alpha\Phi}\mathbf{Tr}\left(
F_{\mu\nu}F^{\mu\nu}\right)  \right]  ,\\
\mathbf{Tr}(.)  &  =\overset{\left(  N-1\right)  (N-2)/2}{\underset{a=1}{%
{\textstyle\sum}
}\left(  .\right)  ,}\nonumber
\end{align}
in which $R$ is the curvature scalar, $\Phi$ is the dilaton field (with
parameter $\alpha)$ and $\mathbf{F}^{\left(  a\right)  }=F_{\mu\nu}^{\left(
a\right)  }dx^{\mu}dx^{\nu}$ stands for the Yang-Mills (YM) field. In terms of
the gauge potentials $\mathbf{A}^{\left(  a\right)  }=A_{\mu}^{\left(
a\right)  }dx^{\mu},$ the YM fields are given by%
\begin{equation}
\mathbf{F}^{\left(  a\right)  }=\mathbf{dA}^{\left(  a\right)  }+\frac
{1}{2\sigma}C_{\left(  b\right)  \left(  c\right)  }^{\left(  a\right)
}\mathbf{A}^{\left(  b\right)  }\wedge\mathbf{A}^{\left(  c\right)  },
\end{equation}
in which $C_{\left(  b\right)  \left(  c\right)  }^{\left(  a\right)  }$
stands for the structure constants of $\frac{\left(  N-1\right)  (N-2)}{2}$
parameter Lie group $G$ and $\sigma$ is a coupling constant. The N-dimensional
YM ansatz is chosen from the Wu-Yang ansatz \cite{R13}%
\begin{align}
\mathbf{A}^{(a)}  &  =\frac{Q}{r^{2}}\left(  x_{i}dx_{j}-x_{j}dx_{i}\right)
,\text{ \ \ }Q=\text{charge, \ }r^{2}=\overset{N-1}{\underset{i=1}{\sum}}%
x_{i}^{2},\\
2  &  \leq j+1\leq i\leq N-1,\text{ \ and \ }1\leq a\leq\left(  N-1\right)
(N-2)/2,\nonumber
\end{align}
where $Q$ is the non-zero YM charge. The YM equations
\begin{equation}
\mathbf{d}\left(  e^{2\alpha\Phi\left(  r\right)  \star}\mathbf{F}^{\left(
a\right)  }\right)  +\frac{1}{\sigma}C_{\left(  b\right)  \left(  c\right)
}^{\left(  a\right)  }e^{2\alpha\Phi\left(  r\right)  }\mathbf{A}^{\left(
b\right)  }\wedge^{\star}\mathbf{F}^{\left(  c\right)  }=0,
\end{equation}
are satisfied by virtue of (5). The corresponding metric functions by setting
$\alpha=\frac{1}{\sqrt{N-2}},$ are found as
\begin{align}
f\left(  r\right)   &  =\frac{\left(  N-3\right)  r}{\left(  N-2\right)
Q^{2}}\left[  1-\left(  \frac{b}{r}\right)  ^{\frac{N-2}{2}}\right]
\nonumber\\
h\left(  r\right)   &  =Q\sqrt{2r}\nonumber\\
e^{2\alpha\Phi}  &  =r.
\end{align}
By following the mass definition for the non-asymptotically flat black holes,
the so-called quasilocal mass introduced by Brown and York \cite{R10}, one can
see that the horizon $b$ is related to the mass $M,$\ charge $Q$ and the
dimensions $N,$ through
\begin{equation}
b=\left[  \frac{2^{\frac{8-N}{2}}MQ^{4-N}}{N-3}\right]  ^{\frac{2}{N-2}}%
\end{equation}

For $b>0$, the horizon at $r=b$\ hides the null singularity at $r=0$\ . On the
other hand, in the extreme case $b=0$ metric (1) still exhibits the features
of the black holes. Because the central singularity $r=0$ is null and
marginally trapped, it prevents outgoing signals to reach external observers.
Using the conventional definition of the statistical Hawking temperature
\cite{R11}, one finds%
\begin{equation}
T_{H}=\frac{1}{4\pi}f^{\prime}\left(  r_{h}\right)  =\frac{\left(  N-3\right)
}{8\pi Q^{2}}.
\end{equation}

One can immediately observe that $T_{H}$ is constant for an arbitrary
dimensions $N$\ and linearly increases with the dimensionality of the
spacetime. As we learnt from black body radiation, radiating objects loose
mass in which the process is governed by Stefan's law \cite{R8}. Therefore
while a black hole radiates, it should also loose mass. According to Stefan's
law, we should calculate the surface area of the black hole (7). The horizon
area $A_{H}$ is found as%

\begin{equation}
A_{H}=\frac{4\pi^{\frac{N-1}{2}}Qb}{\Gamma(\frac{N-1}{2})}%
\end{equation}
where $\Gamma(z)$ stands for the gamma function. After assuming that only
neutral quanta, in respect of the Yang Mills charge, are radiated, Stefan's
law admits the following time-dependent horizon solutions%

\[
b(t)=\exp\left[  -\frac{\sigma}{\left(  2Q\right)  ^{7}\pi^{3}}(t-t_{0}%
)\right]  ,\text{ \ \ \ \ \ \ (}N=4\text{)}%
\]

\begin{equation}
b(t)=\left[  -\mu(t-t_{0})\right]  ^{\frac{2}{N-4}}\text{ ,\ \ \ \ \ \ (}%
N\geqslant5\text{)}%
\end{equation}
where%

\begin{equation}
\mu=2^{-(\frac{12+N}{2})}Q^{-(N+3)}\sigma\frac{\left(  N-4\right)  \left(
N-3\right)  ^{3}}{N-2}\frac{\pi^{(\frac{N-9}{2})}}{\Gamma(\frac{N-1}{2})},
\end{equation}
and $\sigma$\ is Stefan's constant. For $N=4$, the Hawking temperature is
constant with decreasing mass and the black hole reaches to an extreme black
hole state $b=0$ in an infinite time. However, for $N\geqslant5$, the Hawking
temperature has also a constant value for a chosen dimensions $N$, but the
black hole turns out to be the extreme black hole in a finite time according
to $b\sim(t_{0}-t)^{\frac{2}{N-4}}.$ It is observed that the exponential decay
law for $N=4$ dimensions, modifies into a power law decay for $N>4$. Following
the \textit{SCRSM}, we now derive a more precise expression for the
temperature of the black holes (7). To this end, we should first study the
wave scattering in such a spacetime. Contrary to the several black hole cases,
here the massless Klein-Gordon equation
\begin{equation}
\square\Psi=0
\end{equation}
admits an exact solution in the spacetimes (1). The Laplacian operator on the
$N$-dimensional metric (1) is given by%
\begin{equation}
\square=\frac{1}{\sqrt{-g}}\partial_{A}\left(  \sqrt{-g}\partial^{A}\right)
\end{equation}
where $A$ runs from $1$\ to $N.$ One may consider a separable solution as%
\begin{equation}
\Psi=R\left(  r\right)  e^{-i\omega t}Y_{l,m_{1,}..,m_{N-3}}\left(  \theta
_{1},..,\theta_{N-2}\right)
\end{equation}
in which $Y_{l,m_{1,}..,m_{N-3}}\left(  \theta_{1},..,\theta_{N-2}\right)  $
is the spherical harmonics in $N-1$ dimensional space. After substituting
harmonic eigenmodes (15) into the wave equation (13) and making a
straightforward calculation, one obtains the following radial equation:%
\begin{equation}
h(r)^{2-N}\partial_{r}\left[  h(r)^{N-2}f\left(  r\right)  \partial
_{r}\right]  R\left(  r\right)  +\left[  \frac{\omega^{2}}{f\left(  r\right)
}-\frac{l\left(  l+1\right)  }{h(r)^{2}}\right]  R(r)=0.
\end{equation}
After changing the independent variable and the parameters as
\begin{align}
y  &  =1-\left(  \frac{r}{b}\right)  ^{\left(  \frac{N-2}{2}\right)  },\text{
\ \ \ }\tilde{\omega}=\frac{2Q^{2}}{N-3}\omega\\
\tilde{\lambda}^{2}  &  =\frac{2}{\left(  N-2\right)  \left(  N-3\right)
}l\left(  l+1\right) \nonumber
\end{align}
one transforms the radial equation (16) into
\begin{equation}
\partial_{y}\left(  y\left(  y-1\right)  \partial_{y}\right)  R\left(
y\right)  +\left(  \tilde{\omega}^{2}\frac{y-1}{y}-\tilde{\lambda}^{2}\right)
R\left(  y\right)  =0.
\end{equation}
Further, letting%

\begin{equation}
\tilde{\Lambda}^{2}=\tilde{\omega}^{2}-(\tilde{\lambda}^{2}+\frac{1}{4})
\end{equation}
we can obtain the general solution of equation (18) in the form%

\begin{align}
R\left(  y\right)   &  =C_{1}y^{i\tilde{\omega}}F\left(  \frac{1}{2}+i\left(
\tilde{\omega}+\tilde{\Lambda}\right)  ,\frac{1}{2}+i\left(  \tilde{\omega
}-\tilde{\Lambda}\right)  ,1+2i\tilde{\omega};y\right)  +\nonumber\\
&  C_{2}y^{-i\tilde{\omega}}F\left(  \frac{1}{2}-i\left(  \tilde{\omega
}+\tilde{\Lambda}\right)  ,\frac{1}{2}-i\left(  \tilde{\omega}-\tilde{\Lambda
}\right)  ,1-2i\tilde{\omega};y\right)  .
\end{align}
Thus, solution (20) leads to the general solution of Eq. (18) as%
\begin{align}
R\left(  \rho\right)   &  =C_{1}\left(  \frac{\beta-\rho}{\beta}\right)
^{i\tilde{\omega}}F\left(  \frac{1}{2}+i\left(  \tilde{\omega}+\tilde{\Lambda
}\right)  ,\frac{1}{2}+i\left(  \tilde{\omega}-\tilde{\Lambda}\right)
,1+2i\tilde{\omega};\frac{\beta-\rho}{\beta}\right)  +\nonumber\\
&  C_{2}\left(  \frac{\beta-\rho}{\beta}\right)  ^{-i\tilde{\omega}}F\left(
\frac{1}{2}-i\left(  \tilde{\omega}+\tilde{\Lambda}\right)  ,\frac{1}%
{2}-i\left(  \tilde{\omega}-\tilde{\Lambda}\right)  ,1-2i\tilde{\omega}%
;\frac{\beta-\rho}{\beta}\right)
\end{align}
in which
\begin{align}
\rho &  =\left(  r\right)  ^{\left(  \frac{N-2}{2}\right)  }\nonumber\\
\beta &  =\left(  b\right)  ^{\left(  \frac{N-2}{2}\right)  }%
\end{align}
One follows the result of the G. Cl\'{e}ment et al's work (see Eq. (19) in
\cite{R8}), to read the resulting radiation spectrum as
\begin{equation}
\left(  e^{\frac{\omega}{T_{H}}}-1\right)  ^{-1}=\frac{\cosh^{2}\pi\left(
\tilde{\omega}-\tilde{\Lambda}\right)  }{\cosh^{2}\pi\left(  \tilde{\omega
}+\tilde{\Lambda}\right)  -\cosh^{2}\pi\left(  \tilde{\omega}-\tilde{\Lambda
}\right)  }%
\end{equation}
For high frequencies $\tilde{\Lambda}\simeq\tilde{\omega}=\frac{2Q^{2}}%
{N-3}\omega,$ and this leads to
\begin{equation}
e^{\frac{\omega}{T_{H}}}=\cosh^{2}2\pi\tilde{\omega}\rightarrow T_{H}%
=\underset{\omega\rightarrow\text{large value}}{\lim}\frac{\omega}{2\ln\left(
\cosh2\pi\tilde{\omega}\right)  }=\frac{N-3}{8\pi Q^{2}}%
\end{equation}
which is nothing but the statistical Hawking temperature (9).

On the other hand, for $b=0$ (i.e. the case of extreme massless black holes),
the above analysis for computing the Hawking radiation fails. In \cite{R8}, it
is successfully shown that the wave scattering problem in the extreme linear
dilaton black holes in EMD theory reduces to the propagation of eigenmodes of
a free Klein-Gordon field in two-dimensional Minkowski spacetime with an
effective mass. Conclusively, there is no reflection, so that the extreme
linear dilaton black holes can't radiate, although their surface gravities
remain finite. Since setting $b=0$ reduces metric (1) to a conformal product
$M_{2}\times S^{N-2}$ of a two dimensional Minkowski spacetime with the
$\left(  N-2\right)  -$sphere of constant radius, the same interpretation is
valid for the extreme linear dilaton black holes in EYMD theory. Namely, the
massless linear dilaton black holes in EYMD theory also can't radiate.

\section{Einstein-Maxwell-Dilaton (EMD) black hole radiation spectrum}

In order to make a comparison with the foregoing YM results in this section we
represent the similar analysis for the EMD case. The $N-$dimensional EMD
action is given by \cite{R12}%
\begin{equation}
S=\int d^{N}x\sqrt{-g}\left(  R-\frac{4}{N-2}\left(  \nabla\Phi\right)
^{2}-e^{-\frac{4\alpha\Phi}{N-2}}F^{2}\right)
\end{equation}
where $\alpha$ is the dilaton parameter, and $F^{2}=F_{\mu\nu}F^{\mu\nu}$
(this form of action was used by Chan et al\cite{R12}). The metric and field
ansaetze are%
\begin{align}
ds^{2}  &  =-f\left(  r\right)  dt^{2}+\frac{dr^{2}}{f\left(  r\right)
}+h(r)^{2}d\Omega_{N-2}^{2}\\
F  &  =e^{\frac{4a\phi}{N-2}}\frac{q}{h^{N-2}}dt\wedge dr.
\end{align}
In Ref.\cite{R12} the general non-asymptotically flat solutions were reported
in which by setting $\alpha=N-3$ one gets the following explicit form%
\begin{align}
f\left(  r\right)   &  =\frac{4}{\gamma^{2}}\left(  \frac{N-3}{N-2}\right)
^{2}r\left(  1-\left(  \frac{b}{r}\right)  ^{\frac{N-2}{2}}\right)  ,\\
h\left(  r\right)   &  =\gamma\sqrt{r},\text{ \ \ }\gamma=\text{constant,}%
\end{align}
where the radius of event horizon $r_{h}=b,$ is given by
\begin{equation}
b=\left(  \frac{2\left(  N-2\right)  M_{QL}}{\left(  N-3\right)  ^{2}%
\gamma^{N-4}}\right)  ^{\frac{2}{N-2}}%
\end{equation}
and $M_{QL}$ stands for the quasilocal mass of the black hole\cite{R10}. One
may use the usual definition of the Hawking temperature at the radius of event
horizon%
\begin{equation}
T_{H}=\frac{1}{4\pi}f^{\prime}\left(  r_{h}\right)  =\frac{\left(  N-3\right)
^{2}}{2\pi\gamma^{2}\left(  N-2\right)  }.
\end{equation}
Following the same procedure as in the previous section, one can show that,
the radiation spectrum of the EMD black holes are same as EYMD case which is
given by Eq. (23), provided the definitions of $\tilde{\omega}$ and
$\tilde{\lambda}^{2}$ are changed as%
\begin{equation}
\tilde{\lambda}^{2}=\frac{l\left(  l+1\right)  }{\left(  N-3\right)  ^{2}%
},\text{ and }\tilde{\omega}=\frac{\left(  N-2\right)  \omega\gamma^{2}%
}{2\left(  N-3\right)  ^{2}}.
\end{equation}
With these changes, the high frequency limit of radiation spectrum gives%
\begin{equation}
e^{\frac{\omega}{T_{H}}}=\cosh^{2}2\pi\tilde{\omega}\rightarrow T_{H}%
=\underset{\omega\rightarrow\text{large value}}{\lim}\frac{\omega}{2\ln\left(
\cosh2\pi\tilde{\omega}\right)  }=\frac{\left(  N-3\right)  ^{2}}{2\pi
\gamma^{2}\left(  N-2\right)  }%
\end{equation}
which is matched with (31). In Fig. (1) we plot the high frequency limits of
the Hawking temperatures in terms of the dimensions of the space-time, for
both EMD and EYMd cases. Also, in Fig. (2) we plot the Hawking temperature in
terms of the spectrum low frequency $\omega$ for both EMD and EYMD black hole.

\section{Conclusion}

The semi-classical radiation spectrum method (SCRSM) properly works to compute
the Hawking radiation for massive higher-dimensional $(N\geq5)$ linear
dilatonic black holes in EYMD and EMD theories. In the high frequency region,
the results of SCRSM agree with the temperature $(T_{H})$ obtained from the
surface gravity. The dependence of $T_{H}$ on the dimensionality $N$ is
plotted for both the EYMD\ and EMD theories to see the differences. From the
Stefan's law it is shown that the $N=4$ dimensions dilatonic black holes
evaporate in an infinite time, while for $N\geq5$ it takes a finite time. We
verify also, once more that a massless object can't radiate. We do this by
studying a massless Klein-Gordon field in our extreme dilatonic black hole
background for which $T_{H}=0$, whereas the surface gravity is not zero.

\bigskip

\section{Figure Captions}

Figure 1: A plot of the high frequency limits of Hawking temperatures
$T_{H\text{ }}$ versus the dimensions of the space-time. The solid and crossed
curves represent the EMD and EYMD cases, respectively.

Figure 2: $T_{H\text{ }}$versus small frequency plot in 5D. Up to a systematic
shift, the EMD and EYMD both exhibit similar behavior at small frequencies.

\end{document}